\documentclass[journal=jacsat,manuscript=article]{achemso}
\usepackage[utf8]{inputenc}

\usepackage{amsmath}
\usepackage{amsmath,amssymb}
\usepackage{graphicx}
\usepackage{caption}
\usepackage{color}
\usepackage{dcolumn}
\usepackage{bm}
\usepackage{float}
\usepackage{subfig}
\usepackage{soul}
\usepackage{comment}

\usepackage{xr}
\usepackage{rotating}

\SectionNumbersOn

\author{Eric Boittier} \affiliation[University of Basel]{Department
of Chemistry, University of Basel, Klingelbergstrasse 80 , CH-4056
Basel, Switzerland.}

\author{Mike Devereux} \affiliation[University of Basel]{Department
of Chemistry, University of Basel, Klingelbergstrasse 80 , CH-4056
Basel, Switzerland.}

\author{Markus Meuwly} \affiliation[University of Basel]{Department of
  Chemistry, University of Basel, Klingelbergstrasse 80 , CH-4056
  Basel, Switzerland.}  \email{m.meuwly@unibas.ch}

\title{Molecular Dynamics with Conformationally Dependent, Distributed
  Charges}
\date{\today}
\begin{document}

\begin{abstract}
Accounting for geometry-induced changes in the electronic distribution
in molecular simulation is important for capturing effects such as
charge flow, charge anisotropy and polarization. Multipolar force
fields have demonstrated their ability to qualitatively and correctly
represent chemically significant features such as sigma holes. It has
also been shown that off-center point charges offer a compact
alternative with similar accuracy. Here it is demonstrated that
allowing relocation of charges within a minimally distributed charge
model (MDCM) with respect to their reference atoms is a viable route
to capture changes in the molecular charge distribution depending on
geometry. The approach, referred to as ``flexible MDCM'' (fMDCM) is
validated on a number of small molecules and provides accuracies in
the electrostatic potential (ESP) of 0.5 kcal/mol on average compared
with reference data from electronic structure calculations whereas
MDCM and point charges have root mean squared errors of a factor of 2
to 5 higher. In addition, MD simulations in the $NVE$ ensemble using
fMDCM for a box of flexible water molecules with periodic boundary
conditions show a width of 0.1 kcal/mol for the fluctuation around the
mean at 300 K on the 10 ns time scale. The accuracy in capturing the
geometry dependence of the ESP together with the long-time stability
in energy conserving simulations makes fMDCM a promising tool to
introduce advanced electrostatics into atomistic simulations.
\end{abstract}

\section{Introduction}
Electrostatics are key to describing nonbonded interactions between
polar molecules or functional groups. As well as governing the
strength of an interaction via Coulombic attraction or repulsion, the
anisotropy of the electron density - such as lone-pairs or sigma-holes
- governs
directionality,\cite{Auffinger:2004,Lommerse:1996,Hardegger:2011} and
spatial arrangement of polar regions contribute to interaction
specificity in environments such as protein binding
sites\cite{Politzer:1985,Kukic:2010} or ionic and eutectic
liquids.\cite{castner:2011}\\

\noindent
Different approaches have therefore evolved to accurately describe
electrostatic interactions. Due to favourable scaling in the number of
pairwise interactions in the condensed phase, simple point charge (PC)
models that are relatively easy to obtain with interaction terms that
are quick to evaluate are still
prevalent.\cite{charmm27,christen:2005,case:2005} Advances in
computational power have led to increased interest in atomic multipole
expansions as a means of including additional anisotropy at a
moderately increased computational
cost.\cite{Lagardere:2017,Popelier:2015,ponder:2010,MM.Stark:2009}
Distributed charge models (point charges that are placed away from
nuclear positions) offer an efficient alternative to multipole moments
by using Machine Learning (ML) techniques to identify a minimal set
(minimally distributed charge model - MDCM) that describe the electric
field around a molecule to a desired level of
accuracy.\cite{MM.dcm:2014,MM.mdcm:2017,MM.MDCM:2020} A further
alternative, the Gaussian Electrostatic Model (GEM), additionally
offers improved close-range interactions relative to a truncated
multipole expansion.\cite{cisneros:2012}\\

\noindent
These approaches typically apply static electrostatic terms that are
distributed over atomic sites. The terms adapt to conformational
changes only by moving with nuclear positions and, in the case of
distributed charges or multipole moments, via the change in
orientation of their local axes.\\

\noindent
Quantum chemical analysis reveals that the electron distribution
within a molecule is distorted by conformational change in a more
complex fashion than simply translating and rotating a locally frozen
electron density to a new spatial position and
orientation.\cite{Darley:2008} Electron density is instead free to
flow towards one nucleus or away from another upon stretching a bond,
for example, and as a bond cleaves the local electron density may be
profoundly distorted.\\

\noindent
This effect could be corrected for minor structural changes using
averaged electrostatics that attempt to describe the molecular
electric field adequately for a range of conformers using a single,
fixed electrostatic model with charges either located at the position
of the nuclei or away from them. Alternatively, fluctuating charge
models exist that assign nuclear charges dynamically in response to
changes in distance to adjacent atoms based atomic electronegativity
and chemical hardness.\cite{Rick:1994,Jensen:2022,Patel:2004} More
recently, approaches have been developed to describe the distortion of
the molecular electric field with conformational change more
precisely. Piquemal and coworkers fitted multipole moments and
charge-flow terms as a function of water molecule
geometry.\cite{Chengwen:2020} A study of CO in myoglobin employed a
3-site point charge and multipolar model with magnitudes that respond
to bond-length to accurately describe free ligand dynamics within a
protein.\cite{MM.MbCO:2003,MM08mtp} ML approaches have also been
developed to predict multipole moments directly from molecular
geometry\cite{Darley:2008} and were recently used in a dynamics
study.\cite{Symons:2021} However, such approaches are still at their
explorative stage and are yet to be applied to the simulation of
larger, conformationally flexible molecules.\\

\noindent
Here we present an alternative approach based on representing the ESP
itself, whereby distributed charges adapt their positions in response
to changes in molecular conformation. The resulting electrostatic
interactions between point charges retain the advantage of being rapid
to evaluate during molecular dynamics simulations in the condensed
phase, and integrate easily into existing force field frameworks in
place of static charges placed at nuclear positions. As charge
magnitudes are fixed there is also no need for bookkeeping techniques
to maintain the correct total molecular charge. The approach is
applied to water, formic acid, formamide and dimethyl ether and
integrated into the CHARMM molecular dynamics engine to validate
implementation and compatibility with existing simulation tools.\\

\noindent
This work is structured as follows. First, the methods are
described. Next, the quality of the flexible MDCM (fMDCM) is assessed
and compared with results from MDCM and PC representations for the
four molecules. This is followed by MD simulations for flexible water
in small clusters and for bulk using periodic boundary
conditions. Finally, the susceptibility of the fMDCM model to general
perturbations in molecular structures is probed and compared with MDCM
and PC parametrizations, followed by conclusions.\\

\section{Methods}

\subsection{Reference Calculations}
Four molecules - water, formic acid, formamide and dimethyl ether -
were chosen as test cases. These models were kept deliberately simple
to lessen the influence of degrees of freedom not considered on
interpretations regarding charge flow. The geometries were optimized
at the $\omega$B97XD/6-311G(2d,2p) level of theory, using
Gaussian09.\cite{g09} Following the confirmation of zero imaginary
frequencies, relaxed scans along internal degrees of freedom were
performed using the opt=ModRedundant keyword. The angle $\theta$ was
scanned over a range $\pm 20^\circ$ in increments of $1^{\circ}$
around the minimum energy structure and for the bonds the range
covered $\pm 0.1$ \AA\/ around the minimum for 20 steps in each
direction. Only for the OH bond $\pm 0.05$ \AA\/ and 10 steps in both
directions was scanned. Such small increments in internal coordinates
were used to ensure that the change in charge positions between the
points remain smooth and continuous, as discussed below. The
$\omega$B97XD density of the optimized geometries was used to generate
the ESPs by using the CubeGen utility in Gaussian09.\cite{g09}\\

\begin{figure}[h!]
    \centering
    \includegraphics[width=15cm]{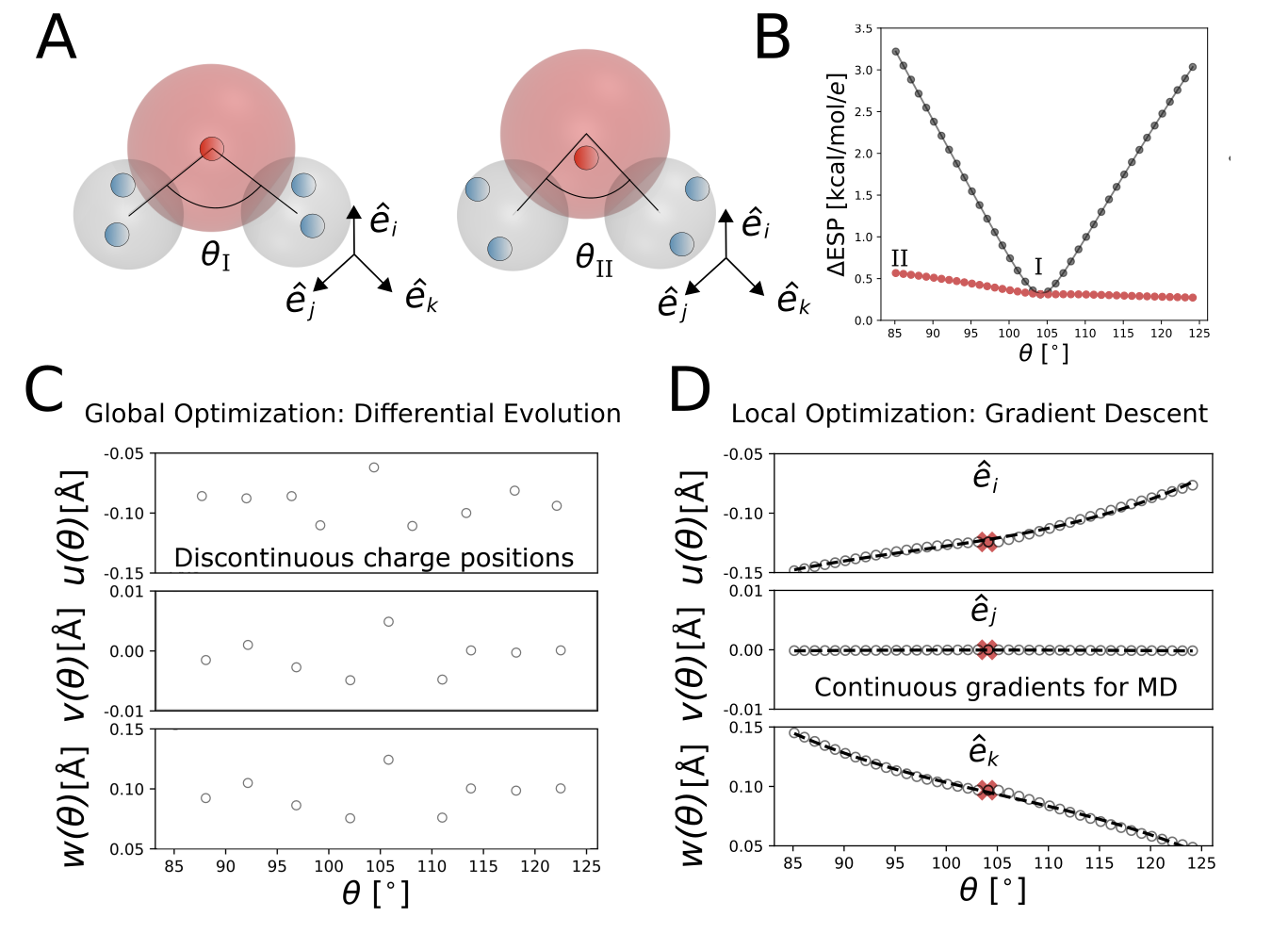}
    \caption{Panel A: Sketch of a 5-charge (small red and blue
      spheres) MDCM model for the equilibrium structure of water
      (structure I with $\theta_{\rm I}$; large red sphere for oxygen,
      large grey spheres for hydrogens) and for a perturbed structure
      (structure II with $\theta_{\rm II}$). Between structures I and
      II, the distributed charge positions adapt. Panel B: quality of
      the ESP from conventional MDCM optimized on the minimum energy
      structure and applied to perturbed structures along $\theta$ for
      water (black circles) and for fMDCM using the MDCM model for the
      minimum energy structure as the reference (red
      circles). Structures I and II are labelled. Panel C: Independent
      fitting of optimal MDCM charges using differential evolution
      leads to discontinuous charge positions ($u(\theta)$,
      $v(\theta)$, and $w(\theta)$) as a function of geometry. Panel
      D: Using gradient descent, starting from the equilibrium
      conformation (red cross) yields continuous charge displacements
      $u(\theta)$, $v(\theta)$, and $w(\theta)$. The vectors
      $\hat{e}_{{\alpha}}$ $(\alpha = i, j, k)$ define the local
      axes.}
    \label{fig:fig1}
\end{figure}

\subsection{Flexible Distributed Charge Model (fMDCM)}
Charge redistribution in fMDCM is captured by introducing a molecular
geometry-dependent position of the off-centered charges, see change in
positions of small blue spheres depending on $\theta$ in Figure
\ref{fig:fig1}A. In the current example a model is developed for
$\theta-$dependent positions of the MDCM charges which leads to a
flexible MDCM (fMDCM) model in which the magnitudes of the
off-centered charges are invariant but their position relative to the
nucleus can vary.\\

\noindent
First, an MDCM charge model of a given order is developed for a
reference geometry which is the equilibrium geometry for all 4
molecules considered. These reference models for water, formic acid,
formamide and dimethylether used 6, 12, 17 and 13 charges,
respectively, which offer a good compromise between accuracy and
computational expense. The positions of the MDCM charges are defined
uniquely relative to a set of local reference frames that are
invariant to molecular translation and rotation, and approximately
retain the charge positions relative to selected neighboring atoms
upon conformational change.\cite{MM.dcm:2014} The reference MDCM model
is determined by minimizing
\begin{equation}
 \Delta {\rm ESP} = \sqrt{\frac{1}{N}
  \sum_{i=1}^{N} (V({\bf r}_i) - V^{\rm{ref}}({\bf r}_i))^2 } 
  \label{eq:loss_fn}
\end{equation}  
where $N$ is the number of ESP grid points used for fitting, typically
$\sim 10^4$ to $10^5$ and specifically 25000 points for water. $V({\bf
  r}_i)$ is the ESP at grid point ${\bf r}_i$ generated by the point
charge model and $V^{\rm ref}({\bf r}_i)$ is the DFT reference value.
Differential evolution (DE)\cite{storn:1997} was used to identify a
minimal set of charges at off-nuclear sites that accurately describe
the ESP around a molecule, typically with an accuracy similar to that
of a multipole expansion truncated at quadrupole or
higher.\cite{MM.mdcm:2017} Symmetry constraints were applied during
fitting to ensure that the fitted charge distributions have the same
symmetry as the parent molecule.\\

\noindent
If independent MDCM models are determined for the $\theta-$perturbed
structures, the MDCM charge positions vary in a non-continuous fashion
due to the stochastic nature of DE, see Figure \ref{fig:fig1}C. This
complicates the interpolation of charge displacements for
$\theta-$values between grid points and also affects the accuracy of
the associated forces. For conceiving a model with smoothly varying
charge positions as the internal coordinates change, gradient descent
$\mathbf{\chi}_{n+1} = \mathbf{\chi}_{n} - \eta \nabla
F(\mathbf{\chi}_{n})$ was used, where $\mathbf{\chi}$ is the vector of
charge positions in the global reference (i.e. $x$, $y$ and $z$) and
$\nabla F(\mathbf{\chi})$ is the gradient of the RMSE in
Eq. \ref{eq:loss_fn} with respect to a change in charge position. A
constant scaling factor $\eta = 0.5$ \AA$^2$ (kcal/mol/$e$)$^{-1}$ was
used to limit the step size in the gradient descent optimization. The
gradient was determined numerically by finite-difference with a step
size of $0.2$ \AA\/. This combination of step size and scaling factor
was found to be appropriate and led to smooth, continuous
displacements of the fMDCM charges as the geometry of the molecule
changes, see Figure \ref{fig:fig1}C.  The sensitivity of $\Delta$ESP
with respect to the number of grid points used to represent the ESP
cube was found to converge to the same error as a fine grid ($6.67$
points/\AA) at a grid spacing of $1.67$ points/\AA. Due to the
significant decrease in computational costs, this grid spacing was
used throughout this study unless mentioned otherwise.  \\

\noindent
As an additional improvement of MDCM itself, conformationally averaged
MDCMs were fitted to multiple conformers by generating a new ESP
reference grid for each conformer, and then transforming each
candidate charge model using its local axes to generate an associated
trial ESP during DE.\\

\begin{figure}[t]
    \centering \includegraphics[width=15cm]{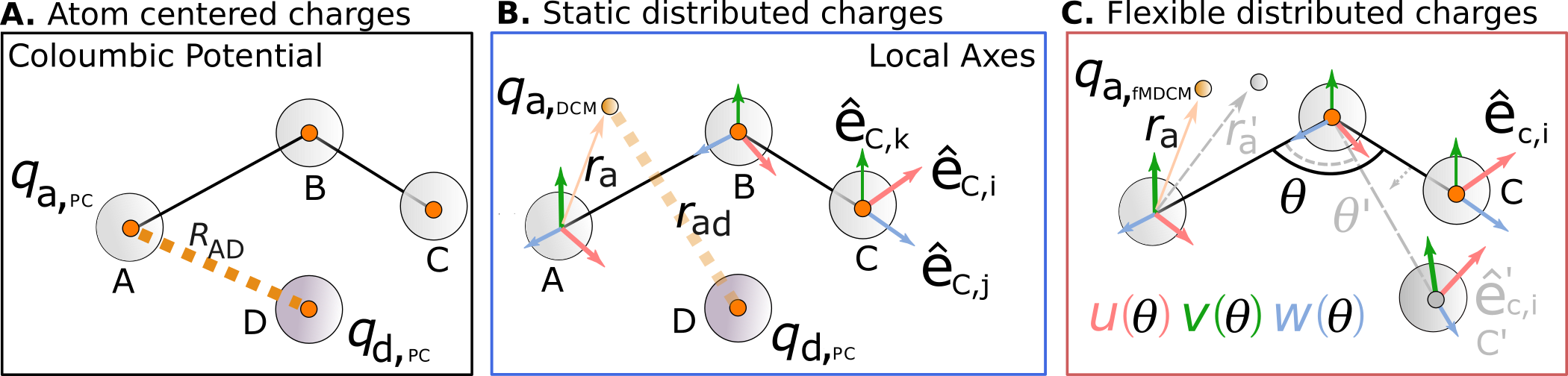}
    \caption{Panel A. The Coulombic interaction and its derivative are
      proportional to the distance between nuclear centers in atom
      centered point charge models. Panel B. In MDCM, this distance
      changes as charges are shifted from nuclear positions within the
      local frame, $\hat{\bf e}$, of three reference axes depicted by
      the red, blue, and green. Panel C. Charge positions within the
      local frame of the fMDCM model also depend on the angle
      $\theta_\mathrm{ABC}$. Parametrized functions $u(\theta)$,
      $v(\theta)$, $w(\theta)$ describe distances along each local
      axis. Variable names set in grey (i.e. $r_a'$, $\theta'$, etc.)
      show how an fMDCM charge responds upon changing $\theta$,
      requiring a modified derivative of the Coulombic potential for
      energy conserving forces in fMDCM.}
    \label{fig:fig2}
\end{figure}

\noindent
For MD simulations the derivatives of the interactions with respect to
the Cartesian coordinates of the nuclei are required. To derive the
necessary expressions, the situation in Figure \ref{fig:fig2} is
considered. Omitting the prefactor $(4\pi \varepsilon_0)^{-1}$, the
Coulomb potential between sites A and D for a conventional point
charge (PC) model is $V \propto \frac{q_{\rm a,PC} \cdot q_{\rm
    d,PC}}{R_{\rm AD}}$, see Figure \ref{fig:fig2}A. The derivative of
the corresponding Coulomb potential with respect to some change in
position of nucleus A, where $\alpha = (x,y,z)$, is then
\begin{equation}
  \frac{\partial V}{\partial R_{{\rm A},\alpha}} = q_{\rm a,PC} \cdot
  q_{\rm d,PC} \frac{\partial }{\partial R_{{\rm A},\alpha}}
  \frac{1}{|{\bf r}_{\rm ad}|}
  \label{eq:DCM_dV}
\end{equation}

\noindent
The situation is similar for distributed charges, but to allow
conformational and rotational transformation of the molecule, ${\bf
  r}_{\rm a}$ is defined relative to a local axis system $\hat{\bf
  e}_{{\rm A,x}}$, $\hat{\bf e}_{{\rm A,y}}$, $\hat{\bf e}_{{\rm
    A,z}}$ defined by atoms $A$, $B$ and $C$, as described
elsewhere.\cite{MM.dcm:2014} Forces on the off-center charge $q_{\rm
  a,DCM}$ of atom $A$ generate torques on $A$, $B$ and $C$ by
re-evaluating equation \eqref{eq:DCM_dV} for these three atoms. The
complete set of partial derivatives for charge $q_{\rm a,DCM}$ in
Figure \ref{fig:fig2}B is:
\begin{align}
\frac{\partial V}{\partial R_{{\rm a},\alpha}} &= \frac{-q_{\rm a,DCM}
  q_{\rm d,PC} \bigl(R_{{\rm
      AD},x}(\hat{\mathbf{\alpha}}\cdot\mathbf{\hat{x}} + g_{1\alpha})
  + R_{{\rm AD},y} (\hat{\mathbf{\alpha}}\cdot\mathbf{\hat{y}}+
  g_{2\alpha})+ R_{{\rm AD},z}
  (\hat{\mathbf{\alpha}}\cdot\mathbf{\hat{z}} +g_{3\alpha})
  \bigr)}{r_{\rm AD}^{3}} \label{Eq:DCMderiv1}\\ \frac{\partial
  V}{\partial R_{{\rm b},\alpha}} &= -\frac{q_{\rm a,DCM} q_{\rm d,PC}
  (R_{{\rm AD},x} g_{4\alpha} + R_{{\rm AD},y} g_{5\alpha} + R_{{\rm
      AD},z} g_{6\alpha})}{r_{\rm
    AD}^{3}} \label{Eq:DCMderiv2}\\ \frac{\partial V}{\partial R_{{\rm
      c},\alpha}} &= -\frac{q_{\rm a,DCM} q_{\rm d,PC} (R_{{\rm AD},x}
  g_{7\alpha} + R_{{\rm AD},y} g_{8\alpha} + R_{{\rm AD},z}
  g_{9\alpha})}{r_{\rm AD}^{3}}
\label{Eq:DCMderiv3}
\end{align}  
where the scalar product $\hat{\mathbf{\alpha}}\cdot\mathbf{\hat{x}}$
is 1 for $\alpha=x$ and zero otherwise. $R_{{\rm AD},x}$ is the
$x$-component of the vector $\mathbf{R}_{\rm AD}$ from nucleus A to
nucleus D. The coefficients $g_{1\alpha}$ to $g_{9\alpha}$ contain the
partial derivatives of the local unit vectors of the frame
($\mathbf{\hat{e}}_{x}, \mathbf{\hat{e}}_{y}, \mathbf{\hat{e}}_{z}$)
with respect to the nuclear coordinates $R_{{\rm a},\alpha}$, $R_{{\rm
    b},\alpha}$ and $R_{{\rm c},\alpha}$. In MDCM the local charge
position is \textit{independent} of $\theta$. The coefficient for the
partial derivatives of frame $\mathbf{\hat{e}}_{x}$ with respect to
atom A, for example, is
\begin{align}
g_{1\alpha} &= u\frac{\partial \mathbf{\hat{e}}_{x,i}}{\partial
  R_{{\rm a},\alpha}} + v\frac{\partial
  \mathbf{\hat{e}}_{y,i}}{\partial R_{{\rm a},\alpha}} +
w\frac{\partial \mathbf{\hat{e}}_{z,i}}{\partial R_{{\rm
      a},\alpha}} \label{Eq:cderiv1}
\end{align}
The prefactors $u$, $v$, and $w$ describe the position of the charge
in the local reference axis system and the remaining coefficients
$g_{2\alpha}$ to $g_{9\alpha}$ have been explicitly given in previous
work.\cite{MM.dcm:2014}\\

\noindent
If atom ``A'' is treated with a fMDCM model, see Figure
\ref{fig:fig2}C, the potential changes to $V \propto \frac{q_{\rm
    a,fMDCM} \cdot q_{\rm d,PC}}{r_{\rm ad}}$ where $r_{\rm ad}$
depends on the A-B-C angle $\theta$ because the position ${\bf r}_{\rm
  a}$ of the fMDCM charge is defined as a function of $\theta$. This
adds a $\theta-$dependence to the vector ${\bf r}_{\rm ad} = ({\bf
  r}_{\rm d} + {\bf R}_{\rm D}) - ({\bf r}_{\rm a} + {\bf R}_{\rm
  A})$.  Component $\alpha$ of ${\bf r}_{\rm a}$ is defined by the
local axes according to
\begin{equation}
  r_{{\rm a},\alpha} = u(\theta) \: {\bf \hat{e}}_{{\rm A,x},\alpha} +
  v(\theta) \: {\bf \hat{e}}_{{\rm A,y},\alpha} + w(\theta) \:
  {\bf \hat{e}}_{{\rm A,z},\alpha}
  \label{eq:DCM_pos}
\end{equation}
Here, $u(\theta)$, $v(\theta)$ and $w(\theta)$ are functions of the
A-B-C angle $\theta$, and describe the distance along $\hat{\bf
  e}_{{\rm A,x}}$, $\hat{\bf e}_{{\rm A,y}}$, and $\hat{\bf e}_{{\rm
    A,z}}$, respectively.\\

\noindent
As fMDCM charge positions are a function of $\theta$, additional terms
enter the derivative for the force evaluations. The partial
derivative, equation \ref{Eq:cderiv1}, becomes
\begin{align}
 g_{1\alpha}(\theta) &= \Bigg[u(\theta)\frac{\partial \mathbf{\hat{e}}_{x,i}}{\partial
    R_{{\rm a},\alpha}} + \mathbf{\hat{e}}_{x,i} \frac{\partial
    u(\theta)}{\partial R_{a,x}} \Bigg] + \Bigg[ v(\theta)\frac{\partial
    \mathbf{\hat{e}}_{y,i}}{\partial R_{{\rm a},\alpha}} +
  \mathbf{\hat{e}}_{y,i} \frac{\partial v(\theta)}{\partial R_{a,x}}\Bigg]
  + \Bigg[ w(\theta)\frac{\partial \mathbf{\hat{e}}_{z,i}}{\partial R_{{\rm
        a},\alpha}} + \mathbf{\hat{e}}_{z,i} \frac{\partial
    w(\theta)}{\partial R_{a,x}}\Bigg]
\label{Eq:cderiv2}
\end{align}
whereby $u(\theta)$, $v(\theta)$ and $w(\theta)$ can be any suitably
parametrized function or numerically defined by using, e.g., a
reproducing kernel.\cite{MM.rkhs:2017} In the present work a cubic
polynomial $u(\theta) = k_1 + k_2\theta + k_3 \theta^2 + k_4 \theta^3$
was used.  The change in local coordinate versus nuclear position is,
for example, $\frac{\partial u(\theta)}{\partial \theta }
\frac{\partial \theta}{\partial R_{a,x}}$. More generally,
$u({\boldsymbol{\rho}})$, $v({\boldsymbol{\rho}})$ and
$w({\boldsymbol{\rho}})$ can be functions of any subset
${\boldsymbol{\rho}}$ of internal coordinates within a molecule that
describe the conformation, such as 4 atoms describing a torsion or
larger sets of atoms describing multiple degrees of freedom. In this
case, the set of partial derivatives is simply extended to also
include these atoms. \\

\noindent
The energy and force expressions were implemented in CHARMM version
c47, and simulations for several test systems were carried out to
illustrate their use, and to verify energy conservation in $NVE$
simulations. The angular dependent terms and associated derivatives
for fMDCM presented here need to be evaluated for each charge at each
simulation time step. This incurs a computational cost which scales
linearly with the number $N$ of charges in the system. For
DCM\cite{MM.dcm:2014} and the same number of charges as for a fixed
point charge model the computational overhead was a factor of $\sim 2$
which, however, has been further reduced in the meantime due to
improvements in the code. As the system size increases, the dominating
factor becomes the charge-charge Coulomb interactions that scale
$\propto N \log N$ and the relative increase in compute time between
fMDCM and PCs with the same $N$ will be considerably smaller than
2. The present implementation of fMDCM also supports parallelization
with MPI and as will be shown below, multi-nanosecond simulations for
water boxes are readily possible.\\

\subsection{Molecular Dynamics Simulations}
A first set of test simulations included one positively charged
potassium ion surrounded by three water molecules described by
fMDCM. The OH bond and HOH angle were flexible with force constants of
$450$ kcal/mol/\AA$^2$ and 55 kcal/mol/rad$^2$ for the stretching and
bending motions, respectively, as available from
CGenFF\cite{MacKerell:2010}. These were used alongside the equilibrium
bond distances and angles ($r_e = 0.9572$\,\AA~and $\theta_e =
104.52^{\circ}$), nuclear charges and Lennard--Jones parameters from
the TIP3P water model.\cite{jorgensen:1983} The simulations started
from a distorted geometry of the cluster using a time step of $\Delta
t = 0.5$ fs and propagating $2 \times 10^6$ time steps in the $NVE$
ensemble. The ion was used to avoid decomposition of the small water
cluster.\\

\noindent
Next, MD simulations with the fMDCM model for water were initiated
from a snapshot of a water box containing $251$ water molecules,
taking coordinates that were previously minimized with CHARMM TIP3P
with SHAKE\cite{Ryckaert:1977} constraints, generated using the CHARMM
GUI server.\cite{jo:2008} The same scheme and force field as for the
previous $NVE$ simulations was used, with $\Delta t = 0.5$ fs. These
simulations employed periodic boundary conditions (PBC) and a $14$ and
$12$\,\AA\/ cut-off for electrostatics and VDW,
respectively. Velocities were initially assigned from a Boltzmann
distribution at $100$\,K; however, no velocity rescaling was carried
out during the $NVE$ simulations. Finally, heating, equilibration and
$NVE$ production MD simulations with PBC were carried out at 300 K for
rigid and flexible water molecules.\\

\noindent
MD simulations were also used to generate structures of the test
molecules to probe the performance of fMDCM models on more generally
distorted structures than specifically scanning along one valence
angle. For generating such perturbed structures for formic acid,
formamide and dimethyl ether, the CGenFF\cite{MacKerell:2010}
parametrization was used whereas for water the same parameters as
previously described were employed. All test molecules were solvated
in a $20$\,\AA$^{3}$ TIP3P water box, equilibrated at $300$ K, with
periodic boundary conditions, using the Nos\'e
thermostat\cite{nose:1984}. All bonds to H-atoms were treated using
SHAKE\cite{Ryckaert:1977}, except for the flexible water
simulations. Coordinates were saved every $200$ fs (formic acid,
formamide, dimethyl ether) or $50$ fs (water). To select a diverse set
of conformers to analyze further, principal component analysis (PCA),
a dimensionality reduction technique, was performed as implemented in
Scipy\cite{SciPy:2020}. PCA aids in interpreting the range of
molecular conformations sampled, which are inherently high dimensional
distributions. The projection was performed using a number of degrees
of freedom, and the input was standardized by subtracting the mean and
dividing by the standard deviation. Dihedral angles ($\phi$) are
transformed by $|\sin{\phi}|$ to recover the isotropic distribution,
before standardization and PCA. The $\Delta$ESP was calculated for
these selected conformers, as described earlier.\\

\section{Results}

\subsection{Quality of fMDCM}
\begin{figure}[htp]
    \centering \includegraphics[width=1\textwidth]{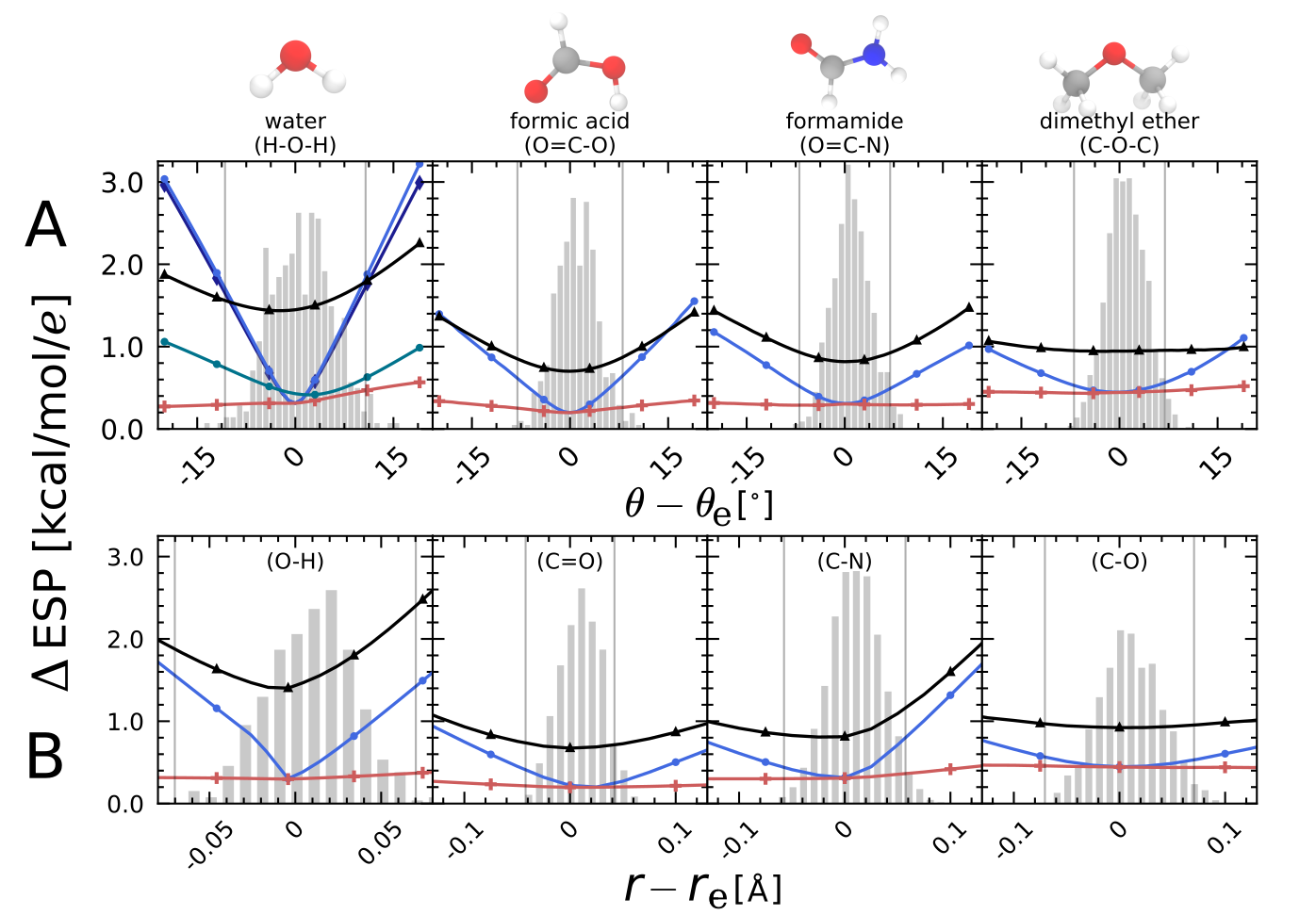}
    \caption{The angle (panel A) and bond length (panel B) dependence
      of the electrostatic potential from the DFT calculations and the
      fitted charge models for point charges (black) and the minimally
      distributed charge models with static (blue) and flexible (red)
      charges. The figure reports $\Delta {\rm ESP}$ as defined in
      Eq. \ref{eq:loss_fn}. The MDCM models for water, formic acid,
      formamide, and dimethylether used 6, 12, 17 and 13 charges,
      respectively. The displacements $\theta - \theta_{\rm e}$ and $r
      - r_e$ are the displacements away from their respective
      equilibrium values $\theta_{\rm e}$. and $r_{\rm e}$. Grey
      histograms represent the approximate distributions of these
      angles which are thermally accessible at 300 K using
      point-charge molecular dynamics. In panel A for water the green
      line is the MDCM model fit to 3 structures and the dark blue
      line is for the ``bisector definition'' of the local reference
      axis system, see also Figure \ref{fig:fig4}. The difference
      between fMDCM and the 3-conformation MDCM model represents the
      amount of charge flux accounted for in fMDCM but not captured in
      MDCM.}
    \label{fig:fig3}
\end{figure}

First, the quality of fMDCM is assessed by determining the accuracy
with which the ESP from electronic structure calculations can be
described for structures generated by scanning along an angle $\theta$
and a selected bond $r$. For water, formic acid, formamide, and
dimethyl ether, the angles considered were the HOH, OCO, NCO, and COC
angles, respectively, whereas the bonds investigated were the OH,
C(sp$^2$)O, CN, and CO (see Figure \ref{fig:fig3}). Three charge
models were explored for each of the four molecules: a conventional PC
model fit to the ESP corresponding to the DFT optimized structure, a
MDCM model optimized for the same structure, and the fMDCM model (as
described in the methods section). PC models were fit to the same ESP
grids as the MDCMs, but least-squares optimization was used in place
of DE. The total charge of the molecule was constrained to zero. For
reference, the $r-$ and $\theta-$ranges covered in finite-temperature
($300$\,K) simulations for each of the compounds in solution are given
as a histogram. The PC models (black symbols and lines) describe the
ESP with typical RMSEs of between 1 and 2 kcal/mol/$e$ for angles
(Figure \ref{fig:fig3}A) and bonds (Figure \ref{fig:fig3}B). The
variation of the difference is within $\sim 0.5$ kcal/mol/$e$ except
for scanning the OH bond in water and the CN bond in formamide for
which the differences are larger.\\

\noindent
Compared with the PC model the best achievable model with MDCM (blue
symbols and lines in Figure \ref{fig:fig3}) is considerably better for
all cases considered. All MDCM models reproduce the reference ESP for
the minimum energy structures to within 0.2 to 0.4 kcal/mol/$e$ which
is a factor of 3 to 5 times better than the best PC model. For
perturbed structures along $\theta$ accessible at ambient conditions
(grey histogram) the maximum RMSE of the MDCM model is comparable to
that from the PC-based model as the rate at which the quality of the
MDCM ESP deteriorates is more rapid when fitted to the equilibrium
conformation only. Fitting to an expanded dataset of several
conformers alleviates this problem, as demonstrated for water: this
model was fitted to three different conformers (equilibrium structure
and distorted geometries corresponding to the solid line in Figure
\ref{fig:fig3}A), leading to a more balanced model (green line) with
similar performance for the equilibrium structure as the
single-conformer MDCM, and similar behavior upon distortion to the PC
model but with lower RMSE. The residual deterioration with
conformational change can be attributed to the use of fixed charge
positions, which do not capture charge redistribution. This result
additionally supports previous findings that fitting multipolar charge
models to multiple conformers separately, then averaging resulting
multipole moments leads to more robust, transferable electrostatic
models.\cite{Kramer:2013} For distortions along the bond $r$ the
changes are also more pronounced but remain below those from the PC
models, see Figure \ref{fig:fig3}B.  \\

\noindent
Finally, the flexible MDCM models (red symbols and lines) perform
uniformly well. All RMSEs are below $\sim 0.5$ kcal/mol/$e$ across the
geometry variations considered and for all cases considered - except
for the scan along $\theta$ for water in Figure \ref{fig:fig3}A - the
difference with respect to the reference ESP is essentially
flat. Hence, capturing the variation of positioning the MDCM charges
while keeping their magnitude constant is a meaningful way to improve
the description of the ESP as a function of internal geometry by a
factor of 2 to 5 when compared with conventional atom-centered point
charges.\\

\begin{figure}[tp]
    \centering \includegraphics[width=1\textwidth]{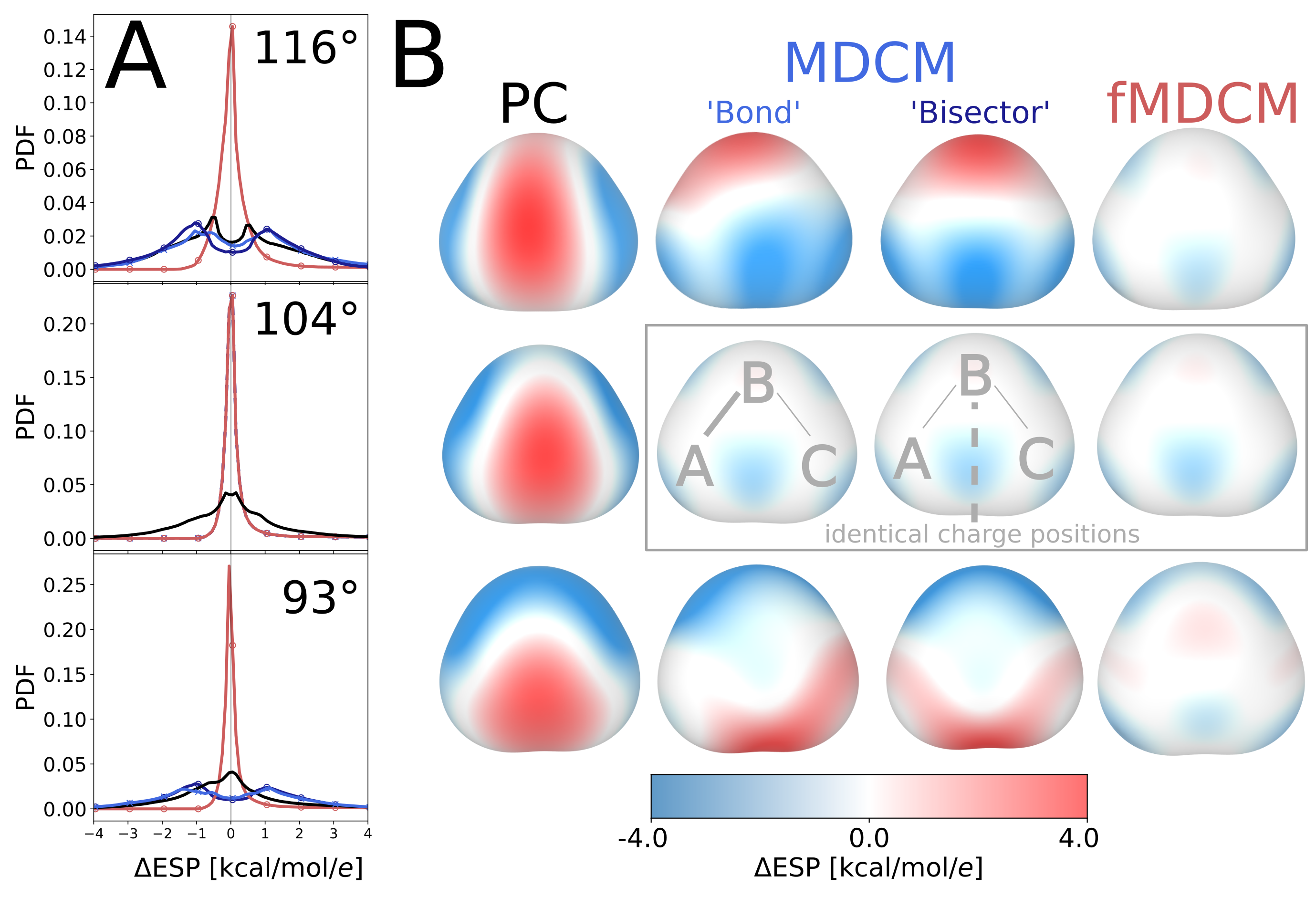}
    \caption{Panel A: The normalized error distributions  between
      DFT ESP and ESP from the different charge models considered:
      point charges (black), MDCM ``bond'' (blue), ``MDCM bisector''
      (dark blue) and flexible distributed charge models (fMDCMs, red)
      for three structures where $\theta = {116^{\circ}, 104^{\circ},
        93^{\circ}}$.  MDCM and fMDCM use 6 charges.  Panel B: Error
      surfaces for all models and geometries. The defining bond (A-B)
      of the local reference system used for distributed charges is
      shown for the equilibrium MDCM surface. Using this bond as the
      local reference breaks the symmetry of the model creating uneven
      error distributions for the non-equilibrium MDCM models. When
      using a ``bisector'' model for MDCM the error distribution
      reflects molecular symmetry for all structures considered.}
    \label{fig:fig4}
\end{figure}

\noindent
Next, the distribution of the RMSEs in the ESP incurred by adopting a
PC, MDCM, and fMDCM representation for water is analyzed. This error
analysis is carried out for DFT densities between $1.0 \cdot 10^{-3}$
and $3.2\cdot10^{-4}$ - representing the local electric field in
regions relevant to MD simulation (Figure \ref{fig:fig4}A) - and in
terms of the location of these errors on the surface of the molecule
at an isodensity value of $10^{-3}$ (Figure
\ref{fig:fig4}B). Conformers with angles of $116^{\circ}$,
$104^{\circ}$, and $93^{\circ}$ were analyzed.  \\

\noindent
The normalized error distributions from the PC model fit to the
equilibrium geometry extends out to $\sim \pm 4$ cal/mol/e, see Figure
\ref{fig:fig4}A (black line). The error projected onto the molecular
0.001 a.u. isodensity surface shows continuous regions of negative
(blue) and positive (red) errors (Figure \ref{fig:fig4}B). Near the
oxygen atom the error surface is typically positive for protracted
angles, and negative for contracted angles, as the magnitude and
positions of the point charges remain constant with respect to nuclear
position and hence do not adequately capture redistribution of the
electron density. As the angle contracts, electron density
(i.e. negative charge) moves from the oxygen atom down into the
bisector of the two OH bonds - creating a region of positive error for
the 93$^{\circ}$ model. Similarly, acute distortions in the valence
angle causes electron density to flow to the oxygen atom. Failure to
account for this flow causes a region of positive error to form on the
oxygen atom for the $116^{\circ}$ PC model. By construction, the point
charge ESP is symmetric. The \textit{ab initio} ESP is also symmetric,
and it follows that the difference between the two is also symmetric
as shown in Figure \ref{fig:fig4}B.\\
    
\noindent
For MDCM two models with different definitions of the local axes were
generated for water. In the first model (``Bond'') the first axis
points along the A-B bond, the second axis is orthogonal to the first
axis and in the plane containing the three atoms and the third axis is
orthogonal to the first and second axis. For such an axis system the
error distribution is asymmetric (see Figure \ref{fig:fig4}A) with
respect to the molecular symmetry as the angle $\theta$ is perturbed
away from the equilibrium geometry (see first and third row of column
``Bond'' in Figure \ref{fig:fig4}B). On the other hand, for the
equilibrium geometry, which was used for the MDCM fit, the error
distribution is manifestly symmetric. If the local axis system uses
the ``bisector'' (third row in Figure \ref{fig:fig4}B) as has been
used in previous work\cite{ponder:2010}, the error distribution
retains the symmetry of the molecular structure as also seen in Figure
\ref{fig:fig4}A. This highlights that the performance of a MDCM or
multipolar electrostatic model upon geometry distortion can be
affected by the choice of local axes, and while in certain cases a
clear choice, such as tying an axis to a dominant bond or bisector,
may be preferable,\cite{MM.bereau:2013} in the general case of lower
symmetry the choice is less clear. The figure also shows that the
average error across all grid points is not typically strongly
affected, however, as has already been observed in Figure
\ref{fig:fig3}A for water (dark vs. light blue lines).\\

\noindent
The error distributions for fMDCM for all three structures are
strongly peaked around zero, as is seen in Figure \ref{fig:fig4}A
(blue line). Also, fMDCM is more robust to the choice of local axes as
charges are free to move back to more optimal positions than the local
axes would have placed them upon distorting a molecule. The projected
error distribution is also considerably more symmetric than for the
corresponding MDCM model, despite using the generic ``Bond'' local
axes. More importantly, using fMDCM greatly reduces the baseline error
of the model, with the maximum of these distributions around $< \pm 1$
cal/mol/$e$ which is about a factor of 5 better than for the PC and
MDCM models for the perturbed structures.\\

\noindent
In summary, Figure \ref{fig:fig4}B highlights the influence of
capturing charge flow on the quality of the ESP determined from the
different point charge-based representations. The particular choice of
local reference axes can add an additional, but minor source of error
for MDCM (or multipolar) models and fMDCM addresses both sources of
errors to provide significant improvement across the entire range of
distorted geometries.\\

\subsection{Molecular Dynamics with fMDCM}
One particularly relevant application for conformationally dependent
charges arises in dynamics studies of solvated compounds. The forces
for MDCM and fMDCM can be determined in closed form and were
implemented in CHARMM for carrying out such simulations. To validate
the implementation, $NVE$ simulations were run for several systems of
increasing complexity. First, three water molecules coordinated to one
positively charged potassium ion in the gas phase were considered, see
Figures \ref{sifig:fig1}A and B. This conveniently obviates the use of
periodic boundary conditions for initial validation. This cluster was
stable for the entirety of the simulation and the coordinates of the
atoms and distributed charges were saved at every time step (1
fs). Unlike the standard MDCM implementation, the distance between
each charge and the atom defining its local reference axis was
flexible, changing smoothly in response to changes in the internal
angle resulting in oscillating displacements of the charge as a
function of time, see Figure \ref{sifig:fig1}C. Figure
\ref{sifig:fig1}D reports the total energy $E(t)$ and its distribution
$P(E)$ for these exploratory simulations.\\

\begin{figure}
    \centering
    \includegraphics[width=1\textwidth]{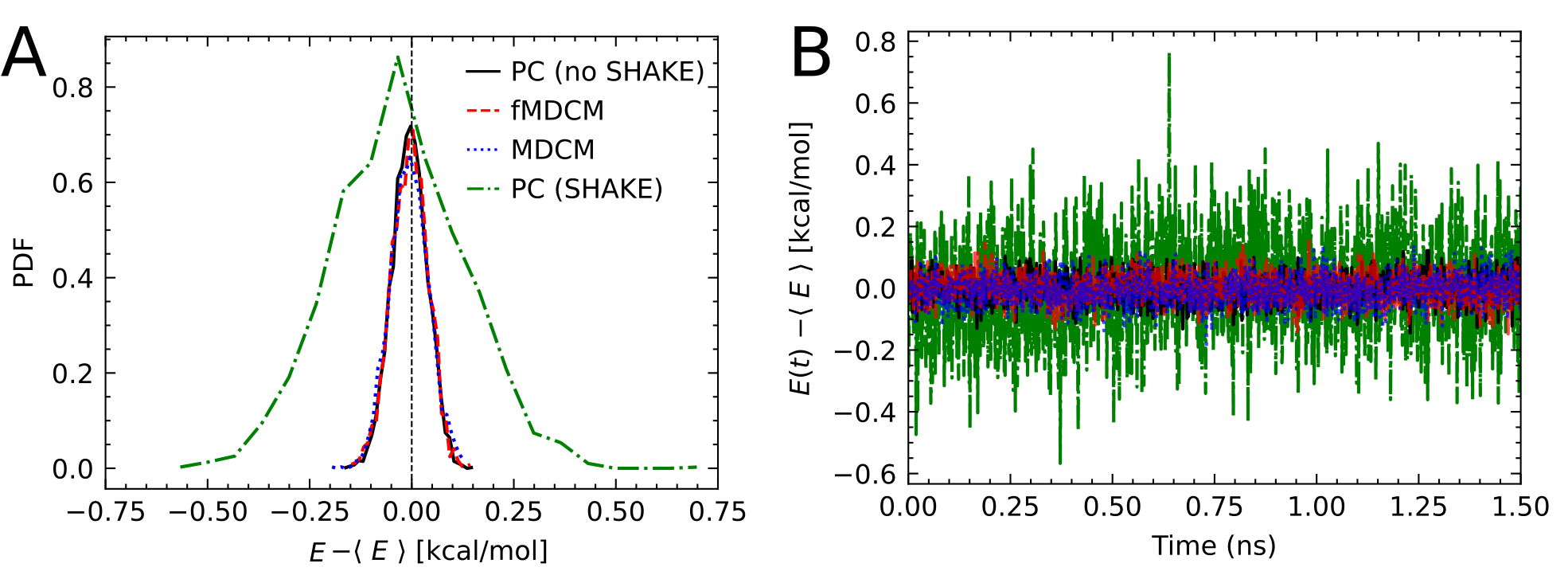}
    \caption{Distribution of energy fluctuations (panel A) and total
      energy time series (panel B) from 1.5 ns simulations at 300 K
      performed with TIP3P (with SHAKE, $\Delta t = 1$ fs, green),
      TIP3P (flexible, $\Delta t = 0.25$ fs, black), MDCM (flexible,
      $\Delta t = 0.25$ fs, red), and fMDCM (flexible, and $\Delta t =
      0.25$ fs, blue). In all simulations total energy is well
      conserved and the distribution of the fluctuation around the
      mean is approximately Gaussian.}
    \label{fig:nve}
\end{figure}

\noindent
A final validation and comparison between different methods was
carried out for a simulation temperature of 300 K. Following heating
(125 ps) and equilibration (125 ps) with periodic boundary conditions,
a 10 ns $NVE$ simulation was run for flexible TIP3P, MDCM, and
fMDCM. The energy fluctuation around the mean for all simulations are
reported in Figures \ref{fig:nve} and \ref{sifig:fig2}. For
simulations with flexible bonds involving hydrogen atoms a shorter
time step needs to be used. Here, $\Delta t = 1$ fs was the time step
for simulations with SHAKE and $\Delta t = 0.25$ fs was used for
simulations with flexible bonds involving hydrogen atoms. For a
flexible water model and the PC, MDCM, and fMDCM models the width of
the Gaussian distributed fluctuation around the mean is $\sim 0.2$
kcal/mol compared with $\sim 0.5$ kcal/mol for TIP3P with SHAKE. Total
energy is manifestly conserved for all four simulations. In addition,
it is of interest to consider the angle time series $\theta(t)$ and
the corresponding distance between one of the 6 conformationally
flexible, off-center charges relative to its defining atom $r_a(t)$,
see Figure \ref{sifig:fig3} (left column). The angle fluctuates by
about $15^\circ$ around the equilibrium value whereas the charge
fluctuates between 0.342 \AA\/ and 0.345 \AA\/ away from the atom it
is defined to. The Fourier transform of the two time series (Figure
\ref{sifig:fig3}, left column) establishes that one mode is related to
the water bending mode (2000 cm$^{-1}$) together with low-frequency
motions (probably due to water-water-water bending) and the bending
overtone at about twice the fundamental frequency are present. For the
same simulations with a time step of $\Delta t = 1$ fs total energy is
also conserved for all models considered, see Figure
\ref{sifig:fig4}.\\

\subsection{General Molecular Deformations}
As a proof of concept the fMDCMs presented so far were fitted to
accurately describe the ESP upon distortion of a single angle, see
Figure \ref{fig:fig3}. The following analyses were geared towards
establishing whether such a model is also suitable to improve the
description of the ESP when more general deformations of the molecules
are allowed, including degrees of freedom that were not accounted for
during fitting.\\

\noindent
The necessary conformations were generated from MD simulations of the
hydrated test molecules which were run at 300\,K. Since the molecular
geometry and ESP depends on $3n$ degrees of freedom, dimensionality
reduction is used to project this space onto a 2D representation. This
is done using principal components (PC1 and PC2), which maximize the
explained variance of the original distribution. The frequency of the
observed MD conformations is proportional to their energy; the peak of
the principal component distribution, centered at the origins in these
projections, corresponds to the equilibrium MD conformer. \\

\noindent
To probe the performance of the model for arbitrarily perturbed
structures, two molecules (formic acid and formamide) were
considered. Based on the results of the relaxed scans in Figure
\ref{fig:fig3}, formic acid appears to be less sensitive to internal
distortions than formamide. Additionally, formic acid requires fewer
internal degrees of freedom to describe its conformation compared to
formamide. To provide an impression of the performance of the model,
formic acid structures were sampled from the parametrized range of
$\theta$, while also sampling a range of $r_2$ bond lengths that were
not part of the parametrization for fMDCM. Formamide conformers were
sampled from the full range of internal angles and dihedrals to more
widely probe the performance of each model across the available
conformational space. The number of selected distorted structures was
kept small (5 for each molecule) for clarity and are intended to
provide guidance as for what regions of conformational space the model
should, and should not, be trusted.\\

\noindent
Five formic acid conformations (a to e) sampled from a simulation in
water were analyzed (Figure \ref{sifig:fig4}). To set the stage,
conformations were selected from the pool of structures such that
$r_1$ was near the equilibrium value and only changes along $\theta$
and $r_2$ occurred. As suggested by the 2D distribution in principal
component space, Figure \ref{sifig:fig4}C, structures a and e are
sampled from the tails of this distribution. For all 5 structures
considered the fMDCM model performs best (Figure \ref{sifig:fig4}D,
red bar), followed by MDCM (blue) and PC (black). However, the
deterioration compared with the performance on the reference structure
(dashed horizontal line) is smallest for the PC model and largest for
fMDCM. Compared with MDCM, the fMDCM model performs better or on par
with it. In other words, the additional boost in accuracy obtained
with fMDCM fitted to a single degree of freedom in this case also
leads to somewhat improved performance for conformers with more
general distortions that are far from the equilibrium conformation and
involve other degrees of freedom, but performance is degraded.\\

\begin{figure}[hp]
    \centering
    \includegraphics[width=1\textwidth]{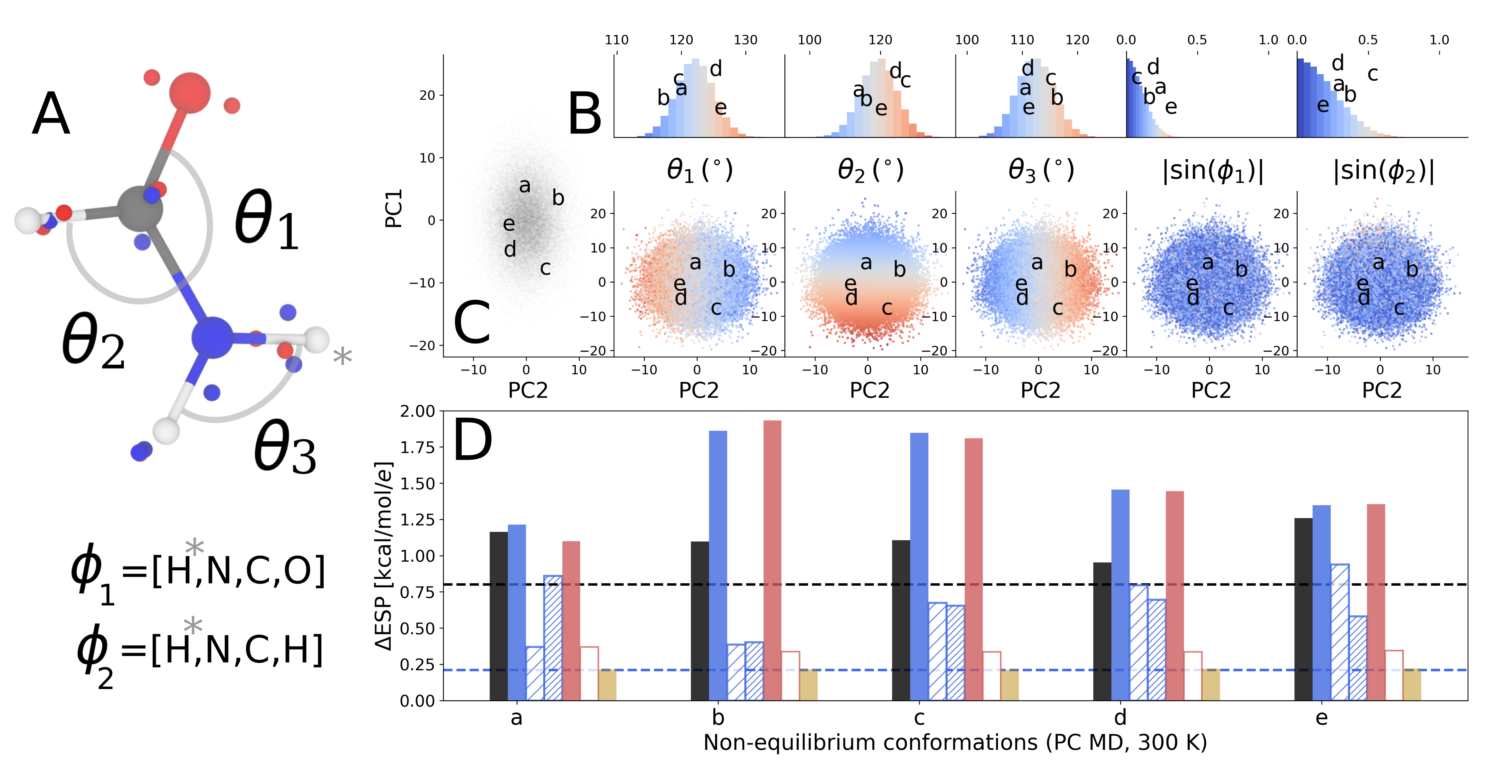}
    \caption{Performance of fMDCM for generally perturbed structures
      of formamide. Panel A: Formamide with 5 select degrees of
      freedom defined. MDCM charges are depicted as red and blue
      spheres. Panel B: 1D distribution of internal coordinates
      sampled from point charge (PC) MD at 300 K. The red-gray-blue
      color scale indicates below-equilibrium, equilibrium, and
      above-equilibrium values, respectively.  Panel C: 2D projection
      of these 5 degrees of freedom in principal component
      space. Color scales are consistent with the 1D distributions
      above. Panel D: RMSEs in the ESP for non-equilibrium geometries
      a to e, comparing PC (black), MDCM (blue), fMDCM for the minimum
      energy structure (solid red), fMDCM for conformers a and b
      (light hatch), fMDCM fit to all five conformers a to e (thick
      hatch), fMDCM for each conformer a to e separately (empty red
      bar), and the 32-pole fit (gold). Dashed, blue and black
      horizontal lines indicate the quality of the equilibrium fit for
      PC and MDCM models, respectively.}
    \label{fig:fig7}
\end{figure}

\noindent
As a second example formamide was considered to probe the quality
outside the parametrized regime, see Figure \ref{fig:fig7}A. The
degrees of freedom considered are angles $\theta_1$ to $\theta_3$ and
dihedrals representing the rotation around the CN bond, $\phi_1$ and
$\phi_2$. The bond lengths involving hydrogen atoms were SHAKEd in the
simulations. Hence, with respect to the DFT equilibrium structures,
the bond lengths are distorted in the samples analyzed.\\

\noindent
As for formic acid, five formamide conformers from the MD simulations
were extracted by selecting geometries with non-equilibrium values of
all degrees of freedom (except for bonds involving hydrogen atoms, see
above) in the tails of the probability distributions (Figures
\ref{fig:fig7}B and C). The ESP from the PC model (black bars in
Figure \ref{fig:fig7}D) differs from the ESP for the minimum energy
structure for which it had been fitted by $\sim 30$ \% and the RMSE is
quite uniform across the five selected structures. This is different
for the MDCM (blue bar) and fMDCM (red bar) models. For those, the
RMSE varies less uniformly and is larger by a factor of 6 to 10
compared with the structure for which they were parametrized (blue
dashed horizontal line). The RMSE can be reduced to comparable levels
as for the reference structure if the MDCM model is optimized for each
of the five structures individually (red open bar) and compares
favourably with a full optimization at the 32-pole level (gold),
suggesting what could be achieved by an fMDCM that depends on all
degrees of freedom simultaneously. This confirms the findings for
point charges as a model that is generally transferable and for
formamide reproduces the ESP uniformly with an accuracy of $\sim 1$
kcal/mol across various conformations. For MDCM it is also confirmed
that the performance can deteriorate for deformations further away
from the reference conformation and for fMDCM if deformations include
regions for which the model was not parametrized.\\

\noindent
As a final comparison, it was investigated to what extent
conformational averaging, as had already been done for water, see
Figure \ref{fig:fig3}A, could reduce sensitivity to conformational
change for formamide - in other words: to what extent transferability
across geometric changes could be retained. Explicitly including ESP
grids of conformers a to e (red hatched bars in Figure
\ref{fig:fig7}D) during DE fitting results in an MDCM model that is on
par or evidently better than PC or MDCMs that do not include such
conformational information. Including the 2 conformers a and b during
fitting leads to a marked improvement of the resulting MDCM when
predicting ESP grids for conformers c to e (see light blue hatch in
\ref{fig:fig7}D), despite their relative distance in conformational
space. This suggests that it is not necessary to widely sample
conformational space in order to improve conformational sensitivity
and transferability of MDCMs. If all five conformers a to e are
included (thick blue hatch in \ref{fig:fig7}D), the overall
performance of MDCM is greatly improved compared with MDCM fit to a
single conformer although except for conformer a, for which the RMSE
is only slightly reduced. An optimal balance in performance can
therefore be achieved by including relevant conformers when fitting
ESP-derived charge models, while significant improvement can already
be obtained by using at least two conformers in order to discard
charge models that are excessively sensitive to conformational
change. In addition, while explicit inclusion of all degrees of
freedom when parametrizing an fMDCM will yield the best results,
fitting the fMDCM from a conformationally-averaged MDCM starting point
should also reduce the performance degradation when sampling
conformational space outside the parametrized range.\\

\section{Discussion and Conclusion}
Distributed charge models are a computationally efficient way to
improve the description of the molecular ESP which is essential to
capture when evaluating electrostatic interactions in quantitative
atomistic simulations. This work demonstrates that a continuous
description for off-center point charges can be obtained using
gradient descent that recovers the conformational dependence of the
reference ESP. The model also yields analytical derivatives with
respect to atom positions that allow energy conserving MD simulations
and the implementation in the CHARMM simulation package was validated
for water treated with periodic boundary conditions.\\

\noindent
For models trained on a single reference structure it is found that
conventional point charges fitted to the ESP have an almost uniform
RMSE (1 to 2 kcal/mol/$e$ for the molecules considered here) for
deformations away from the reference structure. This is different for
MDCM and fMDCM models. For MDCM the RMSE on the fitted structure is
typically lower by a factor of 2 to 5 but increases rapidly for
deformations away from the reference structure. This is much improved
in the case of fMDCM for which the RMSEs remain small for
perturbations along degrees of freedom for which the model was
parametrized. If, however, perturbations of the structures outside the
range or orthogonal to the degrees of freedom for which they were
parametrized occur, the performance can deteriorate appreciably and
even become worse than for a PC-based model. One remedy to this,
explored in the present work, is to develop MDCM models by fitting to
the ESPs of a range of conformers without significant degradation of
the quality of the model of each conformer. Models that are too
sensitive to conformational distortion are thereby discarded during
fitting. Flexible MDCM again offers a robust alternative, however, by
explicitly relaxing charge positions to resolve unfavorable
distortions, but only if all relevant degrees of freedom that describe
the conformational change are accounted for by the model. A future
extension of the fMDCM framework introduced here is to generalize
parametrization of the $u({\boldsymbol{\rho}})$,
$v({\boldsymbol{\rho}})$ and $w({\boldsymbol{\rho}})$ to additional
internal degrees of freedom for which machine learning techniques
offer interesting possibilities.\\

\noindent
The correct description of the ESP with changing molecular geometry
also yields fluctuating molecular dipole moments which can be
advantageous in spectroscopic applications. This has, for example,
been demonstrated from simulations of H$_2$CO with the PhysNet
model.\cite{MM.h2co:2020} PhysNet\cite{MM.physnet:2019} is trained on
energies, forces, and partial charges for a set of nuclear
configurations and therefore includes (atom centered) charge variation
depending on geometry. It was found that for H$_2$CO the relative
intensities of the vibrational bands agree very well with those
observed from experiments. Thus, it will be of interest to use fMDCM
for spectroscopic applications. In addition, the formulation of the
fMDCM model is ideally suited to also incorporate effects of external
polarization. Methods such as the Drude or the ``charge on a spring''
model already use off-center charges to capture
polarization. Incorporating external polarization in fMDCM will amount
to additional slight repositioning of a subset of charges in response
to the external electrical field. \\

\noindent
In addition to charge redistribution, the local axis system used is
found to slightly impact on the error distribution for the ESP upon
conformational change for MDCM. While such issues can be partially
addressed by careful choice of local axes, fMDCM offers a robust
alternative as charge positions are allowed to drift to compensate for
changes in the direction of local axes upon conformational change.\\

\noindent
From the broader perspective of force field development the fMDCM
model provides a starting point for treating nonbonded interactions at
a similar level as compared with kernel- or neural network-based
methods for bonded interactions.\cite{unke:2021,MM.cr:2021} It has
been shown for small molecules that the potential energy surface of a
molecule can be described with exquisite accuracy (${\rm RMSE} \sim
10^{-2}$ to $10^{-3}$ kcal/mol) from using reproducing kernel Hilbert
space\cite{MM.h2co:2020,MM.rkhs:2020} or permutationally invariant
polynomials.\cite{qu:2018} Therefore, combining such ML-based models
for the bonded interactions with fMDCM (and polarizable fMDCM) is
expected to be a powerful extension for condensed-phase
simulations. The remaining term for a comprehensive description of the
inter- and intramolecular interactions, not accounted for so far, are
the van der Waals contributions.\\

\noindent
In conclusion, for the molecules considered here fMDCM is capable of
describing the reference ESP of conformationally distorted structures
on average better by a factor of $\sim 5$ compared with PC and MDCM
while correctly capturing effects of charge rearrangement due to
changes in the molecular geometry. The formulation of a distributed,
flexible point charge-based model presented here generalizes to larger
molecules including deformations along all internal degrees of
freedom. For this, the functions $u({\boldsymbol{\rho}})$,
$v({\boldsymbol{\rho}})$ and $w({\boldsymbol{\rho}})$ need to be
represented either as parametrized expressions or, probably more
conveniently, be learned using a ML-based technique such as a neural
network or a reproducing kernel.\\

\begin{acknowledgement}
This work was supported by the Swiss National Science Foundation
through grant 200020-188724, the NCCR MUST, and the University of
Basel (to MM).
\end{acknowledgement}

\bibliography{refs}

\clearpage

\renewcommand{\thetable}{S\arabic{table}}
\renewcommand{\thefigure}{S\arabic{figure}}
\renewcommand{\thesection}{S\arabic{section}}
\renewcommand{\d}{\text{d}}
\setcounter{figure}{0}  
\setcounter{section}{0}  

\noindent
{\bf SUPPORTING INFORMATION:Molecular Dynamics with Conformationally
  Dependent, Distributed Charges}

\begin{figure}
    \centering \includegraphics[width=0.9\textwidth]{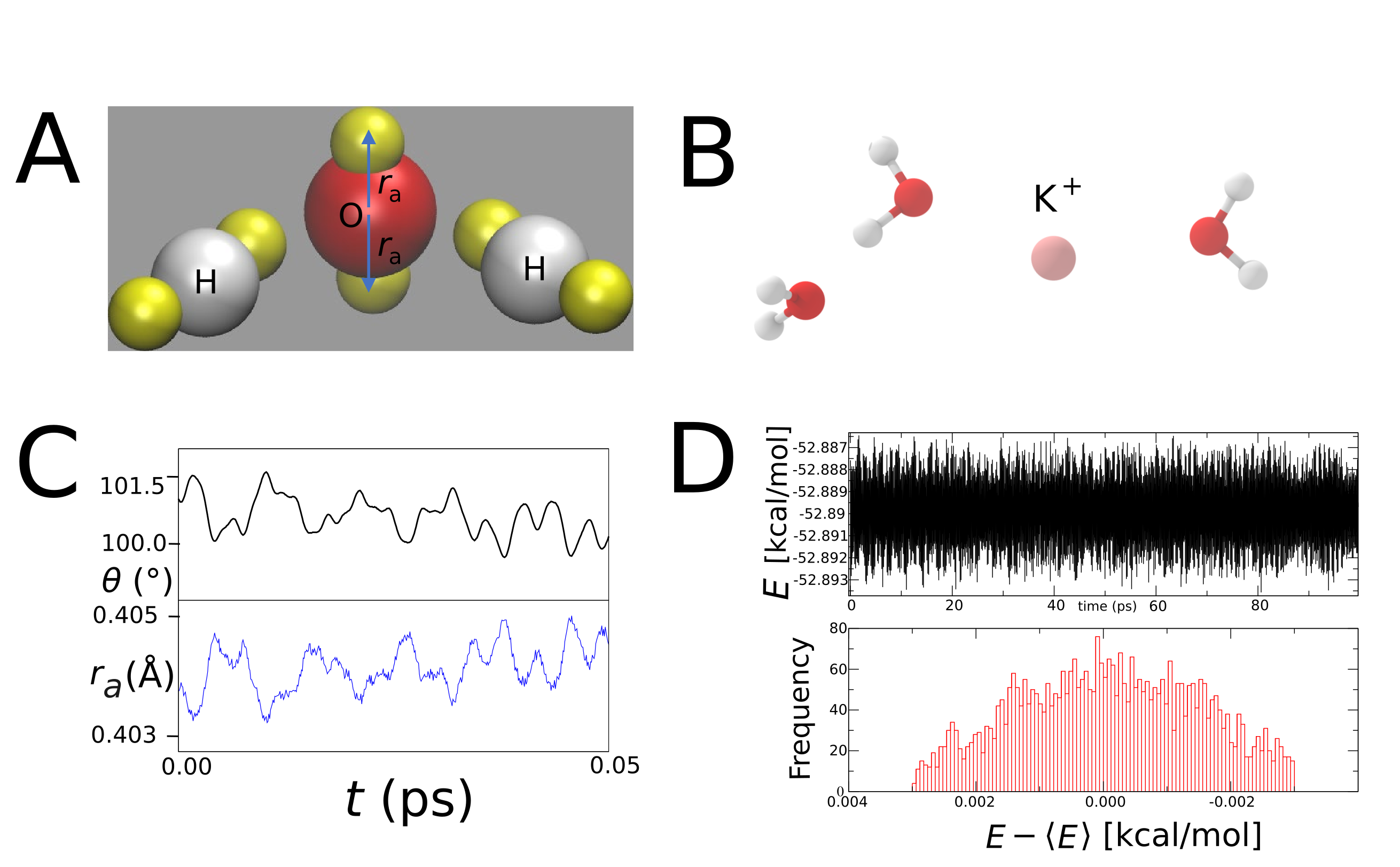}
    \caption{Panel A: The fMDCM water with the 6 off-center charges as
      yellow spheres and the oxygen and hydrogen atoms as red and
      white spheres, respectively, with $\theta$ as the HOH
      angle. Panel B: Explorative simulation system consisting of one
      potassium ion and three fMDCM water molecules. Panel C: Time
      series for angle $\theta(t)$ and the separation of one of the
      fluctuating charges with respect to the atom it is defined to,
      see panel A for definition of $r_{\rm a}$. Panel D: Time series
      for total energy $E(t)$ (top) and histogram for the fluctuation
      around the mean $P(E-<E>)$ (bottom) over 100 ps.}
    \label{sifig:fig1}
\end{figure}

\begin{figure}
    \centering \includegraphics[width=1\textwidth]{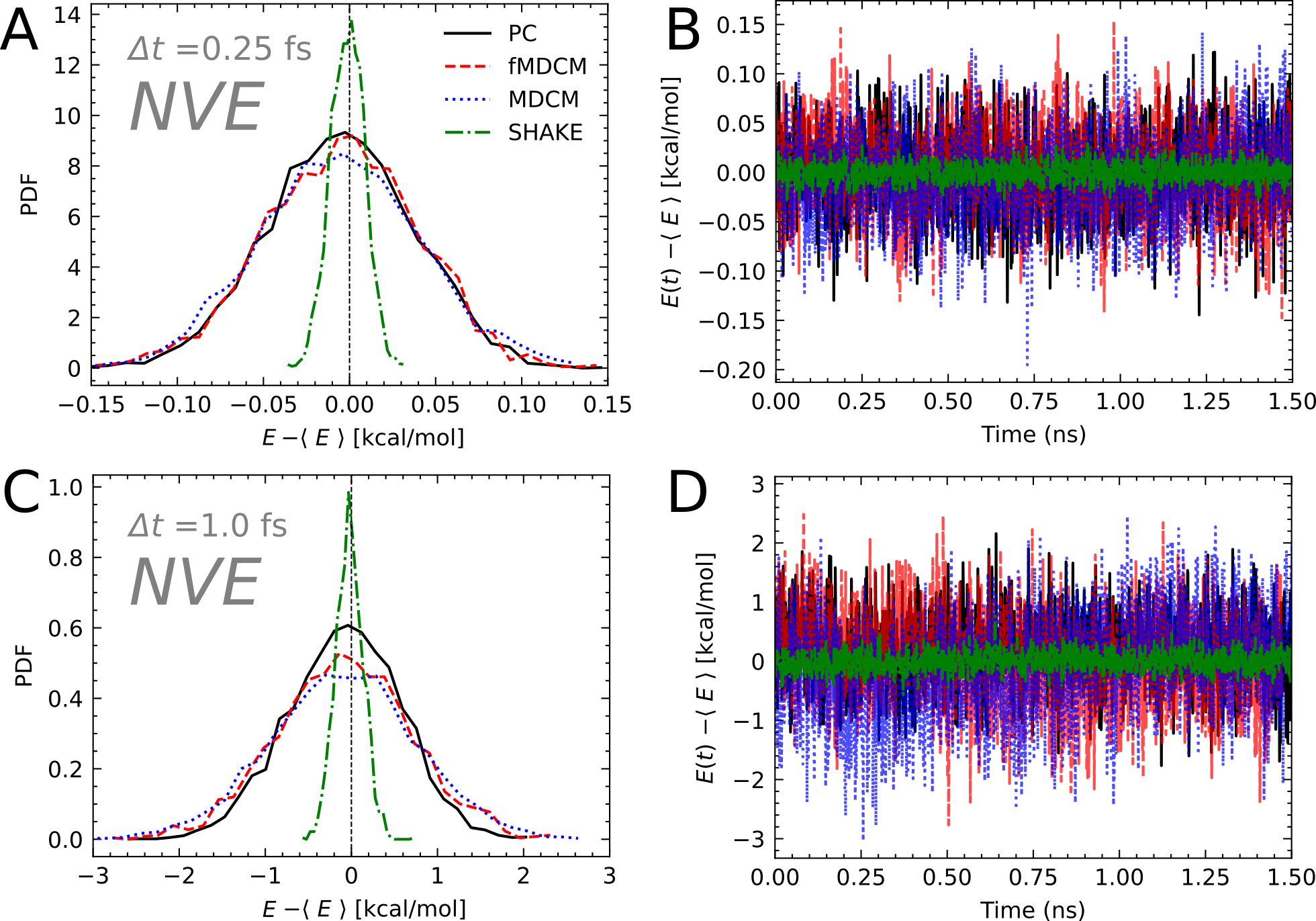}
    \caption{Distribution of energy fluctuations and time series for
      simulations performed at an integration time step of 0.25 fs
      (panels A and B) and 1$\,$fs (panels C and D). Even with
      flexible water for the three electrostatic models total energy
      is conserved for simulations with 1 fs time step and the
      distribution of the energy around the mean is Gaussian. Note the
      different axes ranges for $E -<E>$ in panels A/B vs. C/D.}
    \label{sifig:fig2}
\end{figure}

\begin{figure}
    \centering \includegraphics[width=1\textwidth]{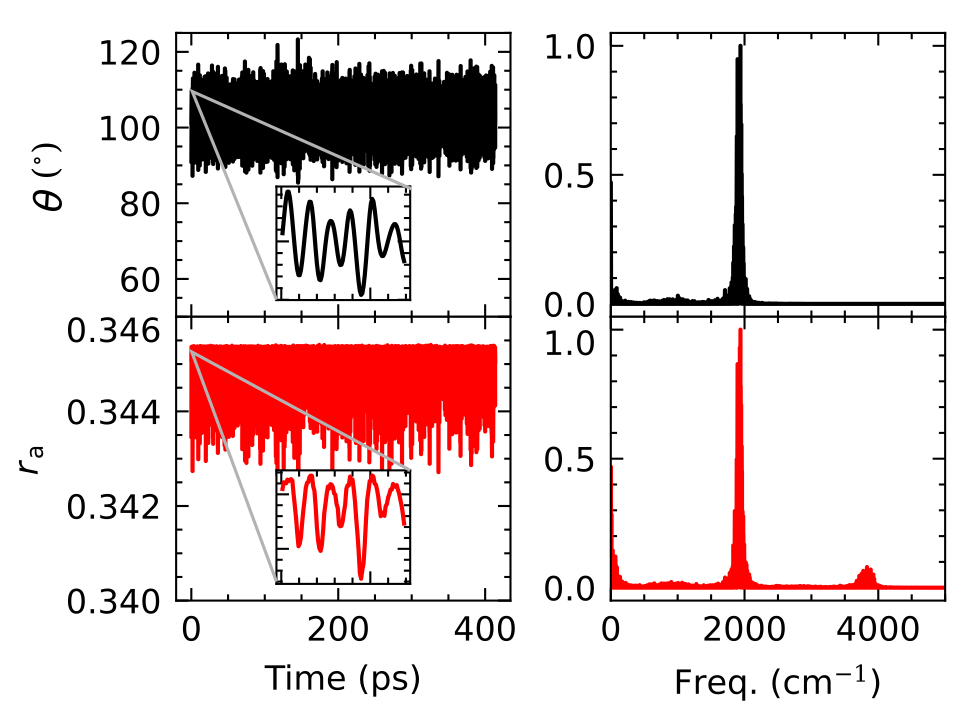}
    \caption{Correlation between valence angle $(\theta)$ and the
      distance between the reference atom (oxygen) and one of the
      associated conformationally flexible charges. The frequency
      spectrum obtained using the fast Fourier transform is also shown
      for the 400 ps trajectory. The simulation was heated and
      equilibrated to 300 K in the $NVT$ ensemble before dynamics were
      collected in the $NVE$ ensemble with $\Delta t = 0.25$ fs.}
    \label{sifig:fig3}
\end{figure}

\begin{figure}[tp]
    \centering \includegraphics[width=0.95\textwidth]{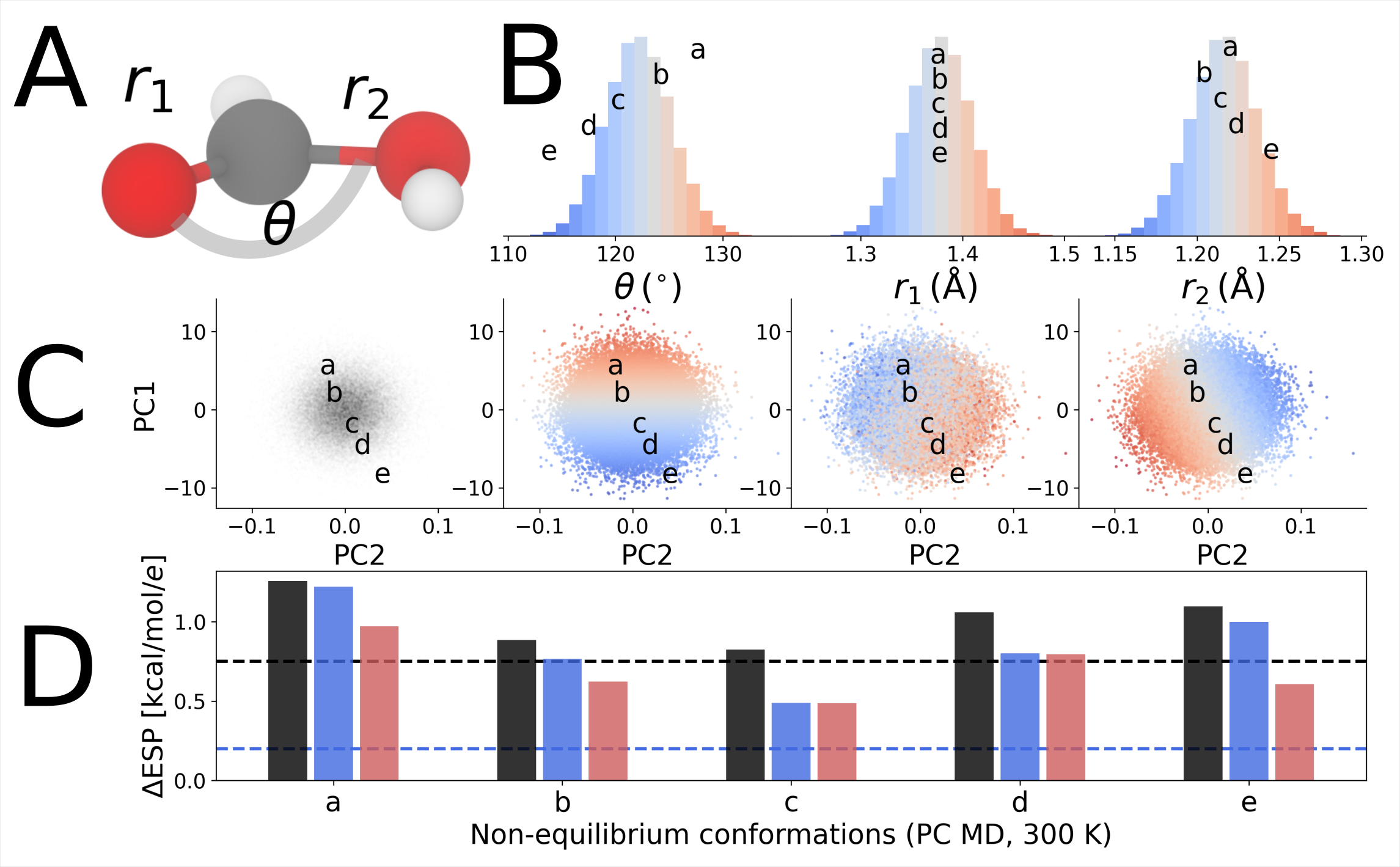}
        \caption{Performance of fMDCM for generally perturbed
          structures of formic acid. Panel A: Formic acid with
          relevant degrees of freedom ($r_1$, $r_2$ and $\theta$)
          shown explicitly. Panel B: 1D distribution of internal
          coordinates sampled from point charge (PC) SHAKE MD at 300
          K.  Panel C: 2D projection of these 3 degrees of freedom in
          principal component space. The red-gray-blue color scale
          indicates below-equilibrium, equilibrium, and
          above-equilibrium values, respectively. Panel D: RMSE for
          the ESP for non-equilibrium geometries, comparing PC
          (black), MDCM (blue) and fMDCM (red) fit to the valance
          angle. Dashed, blue and black horizontal lines indicate the
          quality of the equilibrium fit for PC and MDCM models,
          respectively.}
    \label{sifig:fig4}
\end{figure}

\end{document}